\documentclass[prd,aps,twocolumn,superscriptaddress,amsmath,amssymb]{revtex4}
\usepackage{graphicx,psfrag}
\usepackage{dcolumn}
\usepackage{bm,bbm}

\begin{document}

\title{Anapole dark matter annihilation into photons}

\author{David C. Latimer}

\affiliation{Department of Physics, University of Puget Sound,
Tacoma, WA 98416-1031
}

\newcommand*{\sech}{\mathop{\mathrm{sech}}\limits}
\newcommand*{\balpha}{\boldsymbol{\alpha}} 
\newcommand*{\dilog}{\mathrm{Li}_2}
\newcommand{\qslash}{\not{\hbox{\kern-2pt $q$}}}
\newcommand{\kslash}{\not{\hbox{\kern-2pt $k$}}}
\newcommand{\pslash}{\not{\hbox{\kern-2pt p}}}
\newcommand{\delslash}{\not{\hbox{\kern-3pt $\partial$}}}
\newcommand{\Dslash}{\not{\hbox{\kern-3pt $D$}}}
\newcommand{\gmn}{g^{\mu \nu}}
\newcommand{\Pslash}{\not{\hbox{\kern-2.3pt p}}}
\newcommand{\Kslash}{\not{\hbox{\kern-2.3pt $K$}}}
\newcommand{\Pslashsup}{^\not{\hbox{\kern-0.5pt $^P$}}}
\newcommand{\Poddup}{^\not{\hbox{\kern-0.5pt $^\mathcal{P}$}}}
\newcommand{\Podddown}{_\not{\hbox{\kern-0.5pt $_\mathcal{P}$}}}
\newcommand{\bsig}{\boldsymbol{\sigma}}
\newcommand{\beps}{\boldsymbol{\epsilon}}
\newcommand{\phat}{\hat{\mathbf{p}}}
\newcommand{\khat}{\hat{\mathbf{k}}}
\newcommand{\al}[1]{\begin{align}#1\end{align}}
\newcommand{\diff}{\mathrm{d}}

\begin{abstract}
In models of anapole dark matter (DM), the DM candidate is a Majorana fermion whose primary interaction with standard model (SM) particles is through an anapole coupling to off-shell photons.  As such, at tree-level, anapole DM undergoes p-wave annihilation into SM charged fermions via a virtual photon.  But, generally, Majorana fermions are polarizable, coupling to two real photons.  This fact admits the possibility that anapole DM can annihilate into two photons in an s-wave process.  Using an explicit model, we compute both the tree-level and diphoton contributions to the anapole DM annihilation cross section.  Depending on model parameters, the s-wave process can either rival or be dwarfed by the p-wave contribution to the total annihilation cross section.  Subjecting the model to astrophysical upper bounds on the s-wave annihilation mode, we rule out the model with large s-wave annihilation.  

\end{abstract}

\maketitle

\section{Introduction }

In theories beyond the Standard Model (SM), Majorana fermions are an attractive candidate for the main component of the universe's dark matter (DM).   For example, in supersymmetric theories, if massive neutralinos are the lightest supersymmetric particle, they  can provide the correct thermal relic density for DM if their interaction cross section is at the weak scale \cite{susy_dm}.  
In part, what makes Majorana fermions attractive as DM candidates is their constrained electromagnetic (EM) interactions. 
Because Majorana fermions are self-conjugate fields, they must be electrically neutral with vanishing electric and magnetic dipole moments.  In fact, in terms of static electromagnetic properties, the Majorana fermion can interact only through an anapole moment that couples to virtual (but not real) photons \cite{bk82,nieves,bk83,bk84}. 
From both direct and indirect DM detection experiments, non-observation of DM can be mapped into constraints on their EM interactions.  Early limits on DM dipole moments, polarizabilities,  the anapole moment, and other higher-order interactions can be found in Ref.~\cite{pospelov}, and more recent constraints on possible EM interactions can be  found in Refs.~\cite{Chen:2013gya,Cotta:2012nj,Crivellin:2015wva}.

Despite these constraints, there is ample parameter space to accommodate DM models whose primary interaction is the anapole moment \cite{Fitzpatrick:2010br,anapole_dm1,anapole_dm2,DelNobile:2014eta}.   
In particular, the authors of Refs.~\cite{anapole_dm1,anapole_dm2} propose as a DM candidate a Majorana fermion, $\chi$, that, to leading order, exclusively acts through the anapole moment, $f_a$.  This effective interaction takes the following form in the Lagrangian: $\mathcal{L}_\text{int} = f_a  \overline{\chi} \gamma^\mu \gamma^5 \chi  \partial^\nu F_{\mu \nu}.$
As a consequence, the Majorana fermions do not couple to real photons at tree level but, rather, couple to currents mediated by virtual photons \cite{bk82,nieves,anapole_dm1}.  Dimensional analysis of the Lagrangian term reveals it has a mass dimension of six, so that the anapole moment has mass dimension $[M]^{-2}$.  Supposing the effective interaction is valid below the mass scale $\Lambda$, we can rewrite the anapole moment as  $f_a = g/\Lambda^2$.   Assuming a coupling $g\sim1$  and scale $\Lambda \sim 0.5$ TeV, anapole DM can provide the correct relic density while evading collider constraints for DM masses greater than 100 GeV \cite{anapole_dm2}.   

While a Majorana fermion's anapole moment does not couple to real photons, this does not preclude Majorana fermions from coupling to two real photons.  Indeed, {\em induced} electric and magnetic dipole moments are not generally forbidden for Majorana fermions \cite{radescu}, provided the diphoton process is not jointly odd under parity and even under time reversal \cite{maj_2photon}.    In a UV-complete theory of anapole DM, the Majorana fermion must effectively couple to a charged fermion and scalar or vector boson.  This should generically result in a two-photon coupling via a one-loop box Feynman diagram which, at the Lagrangian level,  results in an effective dimension-7 term $\mathcal{L}_\text{int} \sim \bar{\chi}\chi F^{\mu\nu}F_{\mu\nu}$.  
Constraints on DM interactions through these electric and magnetic polarizabilities can be found in Refs.~\cite{Chen:2013gya,Cotta:2012nj,rayleigh_dm,Frandsen:2012db,Crivellin:2014gpa,Crivellin:2015wva,Ovanesyan:2014fha,Appelquist:2015zfa,Brooke:2016vlw,fichet}.

In a model of anapole DM, two-photon interactions can lead to possible direct or indirect detection signatures for the model. But, perhaps more importantly, the two-photon process admits another possible annihilation channel for DM, which, if appreciable, can significantly impact the relic density of the DM.  
At tree-level, anapole DM can annihilate into two SM charged particles (leptons or quarks below 80 GeV and additionally $W$ bosons and top quarks above this threshold).  The resulting cross section scales as $\Lambda^{-4}$ and is a velocity-suppressed p-wave process \cite{anapole_dm1, anapole_dm2}. The Majorana fermion can also annihilate into two photons via an s-wave process, but power counting suggests additional suppression by the mass scale $\Lambda$.  Because the two-photon effective interaction is described by a dimension-7 term, one would expect the annihilation cross section to scale as $\Lambda^{-6}$, though this could be further suppressed if the model involves an approximate shift symmetry \cite{fichet}.  

Because of the additional scale suppression of this s-wave process, one might anticipate that the p-wave annihilation into SM fermions dominates.   In this paper, we assess, in a model-dependent manner, whether the $\Lambda^{-6}$ suppression of the two-photon contribution to the annihilation cross section is correct.  We assume that the anapole moment is generated through a parity violating coupling between the Majorana DM particle and a charged scalar and fermion.  Then, we compute the s-wave two-photon annihilation cross section that proceeds through box Feynman diagrams.  We find that the scale suppression is predominantly determined by the mass of the charged fermion.  If this charged fermion is the dominant mass in the theory, then the cross section is suppressed by $\Lambda^{-8}$ resulting in a negligible overall contribution; however, if the charged fermion is relatively light, then the two-photon annihilation cross section is surprisingly large, commensurate in size with the p-wave annihilation modes.  In the latter case, this annihilation channel can substantially affect the determination of the DM relic density.  For energies under 80 GeV, we compute the impact of including this process in the total annihilation cross section.

\section{Annihilation amplitude}

In a UV-complete theory for anapole DM, the Majorana fermion will effectively couple to a charged fermion and either a scalar or vector boson in a parity violating manner.  Herein, we only consider the case in which the Majorana fermion of mass $m_\chi$ couples to a scalar particle of mass $m_\phi$ and charged fermion of mass $m_f$; the relevant interaction term in the Lagrangian is
\begin{equation}
\mathcal{L}_\text{I} =  \overline{\psi}(g_{L}P_L + g_R P_R) \chi  \phi^*+ \mathrm{h.c.},  \label{L_int}
\end{equation}
with $P_{R,L}$ right- and left-hand projection operators and $g_R\ne g_L$ assumed.  For a massive thermal DM relic, the DM will be non-relativistic when it drops out of thermal equilibrium, so the relative speed between the annihilating DM particles will be small.  As such, the dominant contributions to the  annihilation amplitude will be the s-wave process of annihilation into two photons and the p-wave process in which the DM annihilates into SM charged particles.  
\begin{figure}[h]
\includegraphics[width=5cm]{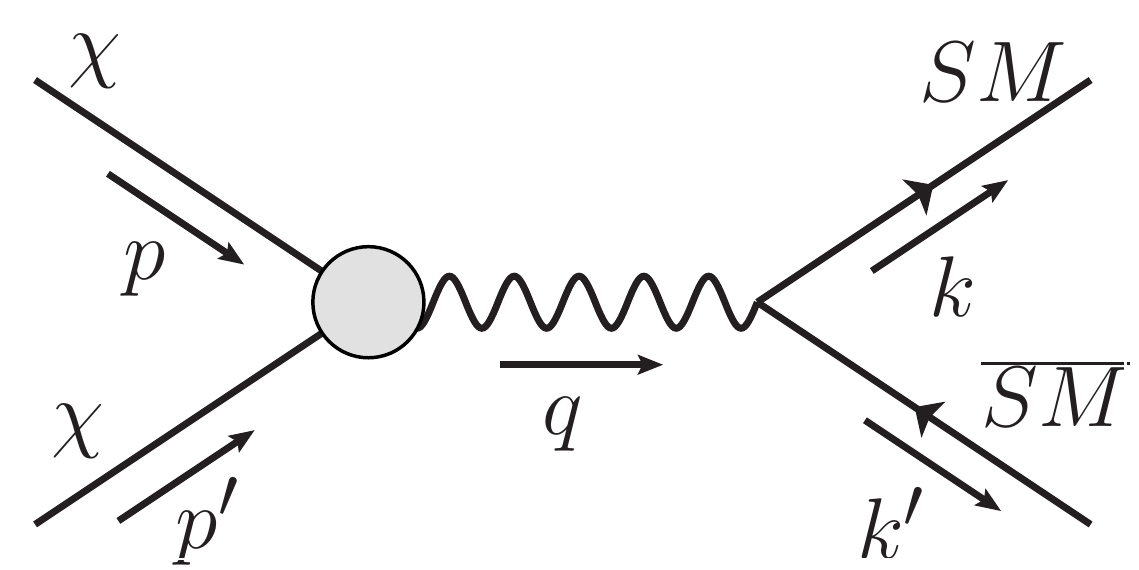}
\caption{ Feynman diagram representing the annihilation of two Majorana fermions into charged SM fermions.    \label{fig1}}
\end{figure}

The p-wave process has been previously considered in Refs.~\cite{anapole_dm1,anapole_dm2}.   As in Ref.~\cite{anapole_dm1}, we restrict the DM mass to lie below the threshold for $W$-boson production, $m_\chi <$ 80 GeV, so that the final state contains only charged SM fermions.  The relevant Feynman diagram for the process is in Fig.~\ref{fig1}; here, the effective anapole vertex entails a factor of $-if_a(q^2 \gamma^\mu - q^\mu \qslash)$ where the photon's momentum is $q = p+p'$.
We  confirm the calculation of the amplitude for this process in Ref.~\cite{anapole_dm1} up to a factor of $i$
\begin{equation}
\mathcal{M}_\text{p} = e f_a  \bar{v}(p') \gamma^\mu \gamma^5 u(p) \, \bar{u}(k) \gamma_\mu v(k').
\end{equation}
In computing the annihilation cross section for unpolarized, non-relativistic DM, we will work in the center-of-momentum (CoM) frame and assume that the final state fermion masses are negligible compared to $m_\chi$.  In this limit, we confirm that this annihilation channel is p-wave with a squared amplitude  in agreement with that in Ref.~\cite{anapole_dm1}
\begin{equation}
|\mathcal{M}_\text{p}|^2 \approx 4 f_a^2 e^2 m_\chi^4  v_\text{rel}^2 (1+ \cos^2\theta) ,
\end{equation}
where $v_\text{rel}$ is the relative velocity between the Majorana fermions, the scattering angle is $\cos \theta = \hat{\mathbf{p}} \cdot \hat{\mathbf{k}}$, and we sum over final spin states.

\begin{figure}[h]
\includegraphics[width=8.6cm]{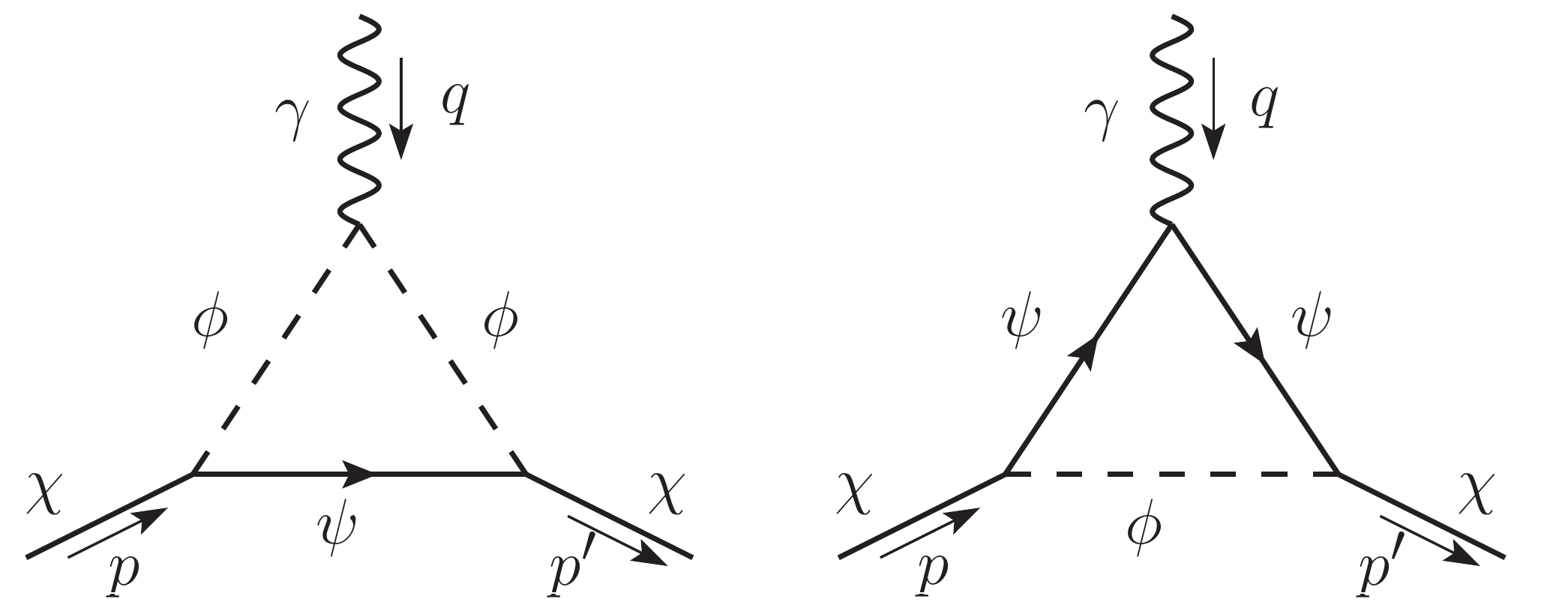}
\caption{ Feynman diagrams relevant for the computation of the anapole moment of the Majorana fermion.  In our computations, we consider both directions of fermion flow in the loop.  \label{fig2}}
\end{figure}
The only remaining computation is that of the model-dependent DM anapole moment.  
 This  involves evaluating the Feynman diagrams in Fig.~\ref{fig2} for an off-shell photon carrying momentum $q^\mu$. The process has the structure $\mathcal{M}^\mu = f_a(q^2) \bar{u}(p')\left( q^2 \gamma^\mu -  q^\mu \qslash \right) \gamma^5 u(p)$, and
the anapole moment $f_a$ is the form factor evaluated at $q^2=0$.  We leave the details to Ref.~\cite{maj_2photon}, and instead, here, only consider two limiting cases.  

Because we  view the anapole vertex as if it were an effective interaction, relevant below some large mass scale, it follows that  either one or both of the loop-particle masses dominate.  In this case, the details of the one-loop processes in Fig.~\ref{fig2} are not appreciable, and the relevant mass scale for the effective interaction is set by the dominant-mass particle in the loop.  As such, we consider two scenarios:  (i) dominant scalar mass, $m_\phi \gg  m_f, m_\chi$ and (ii) dominant charged fermion mass, $m_f \gg m_\phi, m_\chi$.  
 Keeping only the leading order terms in a Taylor expansion of the anapole moment, we find
\al{
{\text{ (i)}}\quad  f_a \approx & \frac{e(|g_L|^2-|g_R|^2)}{(8\pi)^2m_\phi^2} \left[ \frac{4}{3} \log\left( \frac{m_\phi^2}{m_f^2}  \right)- 2 \right],\nonumber\\
& \text{for }\, m_\phi \gg m_f, m_\chi;  \label{ana_i}\\
{\text{ (ii)}}\quad f_a \approx & \frac{e(|g_L|^2-|g_R|^2)}{(8\pi)^2m_f^2} \left[ -\frac{2}{3} \log\left( \frac{m_f^2}{m_\phi^2}  \right)+2\right], \nonumber \\
&\text{for }\, m_f \gg m_\phi, m_\chi. \label{ana_ii}
}
As expected, the anapole moment scales as $f_a \sim \frac{1}{M^2}$ where $M$ is the dominant mass in the loop, and  we also find a subleading logarithmic enhancement of the anapole moment dependent on the ratio of the masses of the charged particles in the loop.  Additionally, we note that the structure of this anapole moment is similar to the computation of the neutrino charge radius in Ref.~\cite{Bernabeu:2000hf}.

\begin{figure}[h]
\includegraphics[width=8.6cm]{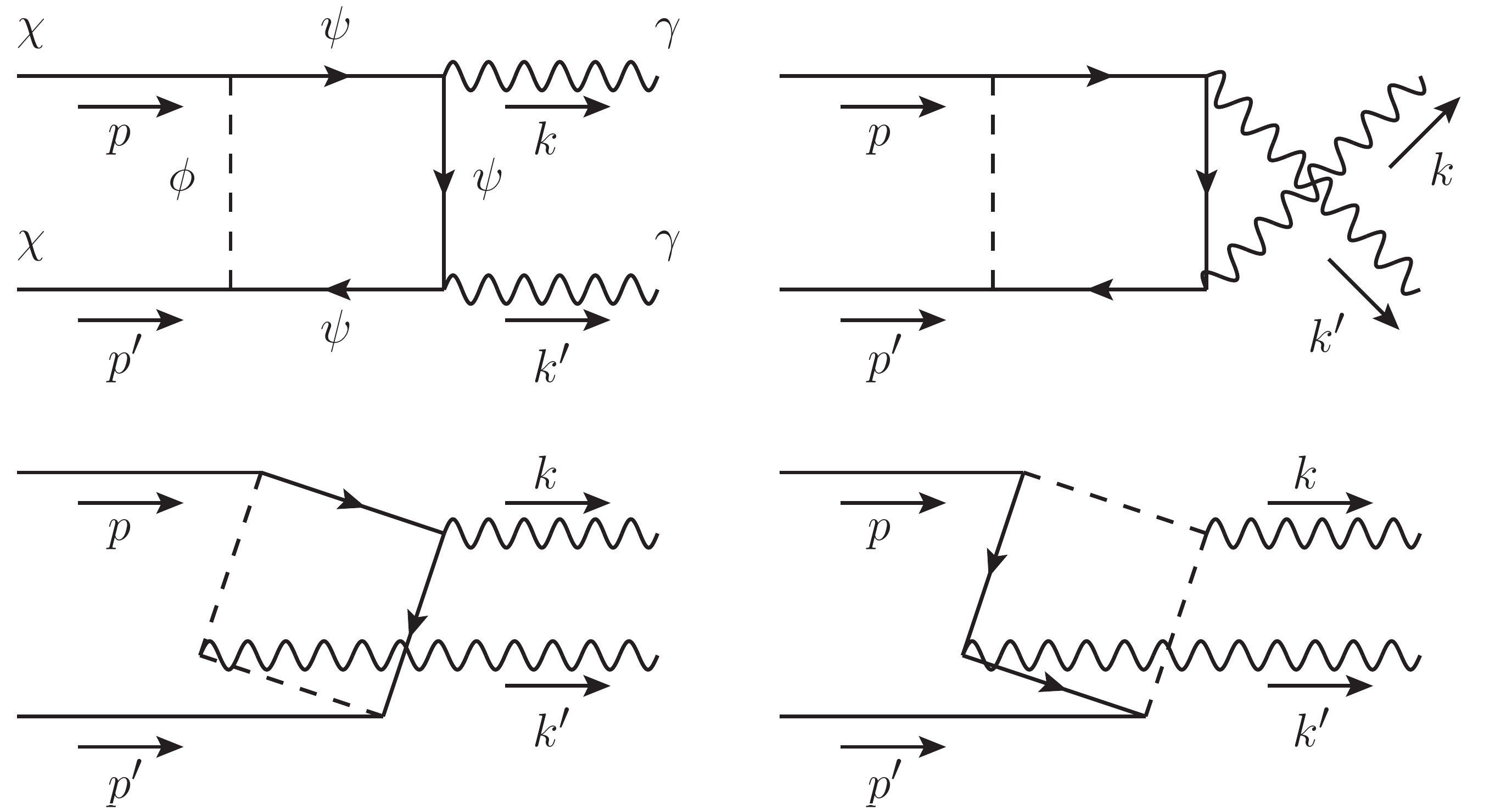}
\caption{Feynman diagrams relevant for s-wave Majorana fermion annihilation into two photons.  In our computations, we consider both directions of fermion flow in the loop.   \label{fig3}}
\end{figure}

We now consider annihilation into two photons.  The relevant Feynman diagrams are in Fig.~\ref{fig3}, and we compute the amplitude for annihilation of two Majorana fermions of opposite spin into two photons in the CoM frame.  Because we are interested in the  s-wave contribution, the Majorana fermions are considered to be at rest $p = p' = (m_\chi, \mathbf{0})$, and the final-state photons have momenta $k=m_\chi (1, \hat{\mathbf{k}})$ and  $k' =m_\chi (1, -\hat{\mathbf{k}})$ with polarizations $\epsilon$ and $\epsilon'$ respectively.   The photons are real and transverse, $k^2= k'^2=0$ and $\epsilon\cdot k = \epsilon'\cdot k'=0$ and have orthogonal polarizations $\epsilon' \cdot \epsilon=0$.    The usual Feynman rules for Dirac fermions must be modified to account for the self-conjugate nature of Majorana fermions; we use the adaptation developed in Refs.~\cite{denner1,denner2}.  

A few remarks are in order on how we effect our calculation. On the face of it, the loop integrals for the diagrams in Fig.~\ref{fig3}  appear to involve four-point functions; however,   partial fraction decomposition can reduce all integrands to, at most, three-point functions.  
 To simplify the expressions for the amplitude, we make ample use of the anticommutation relations for the Dirac gamma matrices along with the Dirac equation for the spinors $u(p), v(p')$.  We use the symbolic manipulation program FORM \cite{form} to aid in these algebraic manipulations. In the end, our expressions involve at most a product of three gamma matrices, but these can be simplified with the identity
 \begin{equation}
 \gamma^\mu \gamma^\nu \gamma^\rho = g^{\mu \nu}\gamma^\rho - g^{\mu \rho} \gamma^\nu + g^{\nu \rho} \gamma^\mu -i \epsilon^{\mu \nu \rho \sigma}\gamma^5 \gamma_\sigma,
 \end{equation}
assuming the convention $\epsilon^{0123} = +1$.  Because we work in the CoM frame and assume s-wave annihilation, the remaining Dirac bilinears can be easily evaluated through explicit computation.  Finally, we remark that, in general, the annihilation amplitude involves two more classes of Feynman diagrams not shown in Fig.~\ref{fig3} -- one with one fermion and three scalar propagators  and the other a seagull diagram; however, these diagrams vanish in the limit of s-wave annihilation.

We find the s-wave component of the amplitude for Majorana fermion annihilation into two photons to be  
\al{
 \mathcal{M}_\text{s} =&   -\frac{i e^2}{8\pi^2}  [\hat{\mathbf{k}}\cdot (\boldsymbol{\epsilon}\times \boldsymbol{\epsilon}')]   \Bigg\{ (  |g_R|^2+ |g_L|^2)        \nonumber\\
 & \times \Bigg[\frac{ m_f^2(I_1 + 2I_3)  }{  m_\phi^2 - m_f^2  + m_\chi^2} +  \frac{ m_\phi^2 I_2- m_f^2 I_3 }{m_\phi^2 - m_f^2} \Bigg ] \nonumber \\
&+ 2 \mathrm{Re} [g_R^*g_L]   \frac{ m_f m_\chi(I_1 +2I_3 )} {m_\phi^2 - m_f^2 +m_\chi^2} \Bigg\} ,
}
with the integrals, $I_j$, defined below.  
Expressing the integrals in terms of the mass ratios $a:= \frac{m_\chi^2}{m_\phi^2}$ and $b:= \frac{m_f^2}{m_\phi^2}$, we have
\al{
I_1 :=& \int_0^1 \frac{1}{x}\log\left( \frac{4 a x^2 -4 a x + b}{b} \right) \diff x, \label{I1} \\
I_2 :=& \int_0^1 \frac{1}{x}\log\left( \frac{-a x^2 +(b+a -1)x +1}{a x^2 +(b-a - 1)x +1} \right) \diff x, \label{I2}\\
I_3:=& \int_0^1 \frac{1}{x}\log\left( \frac{-a x^2 +(1 +a - b)x +b}{a x^2 +(1 -a -b)x +b} \right) \diff x. \label{I3}
}
We note that this computation of the amplitude is similar to one contained in Ref.~\cite{bu}.  Our results are consistent with those in Ref.~\cite{bu} aside from the fact that the authors in that work make use of a projection operator to select out the s-wave contribution to neutralino annihilation into photons.

 As with the anpapole moment, we focus only upon  limiting cases, refinements of those considered above:   (i.a) $m_\phi \gg  m_\chi \gg m_f$, (i.b) $m_\phi \gg m_f \gg m_\chi$,  and (ii) $m_f \gg m_\phi \gg m_\chi$.  We discuss our approximations of the integrals, Eqs.~(\ref{I1} -- \ref{I3}), in the Appendix.
 The first limit, (i.a), is relevant if the DM candidate and charged scalar are beyond-SM (BSM) particles, but the charged fermion is a SM particle.  
 Cases (i.b) and (ii) are relevant if the charged particles are BSM with masses greater than the DM candidate.  If we assume the existence of a discrete symmetry like R-parity, then the DM candidate, as the lightest BSM particle, would be stable to decay for all these cases.
 We make one further simplification; that is, we assume maximal parity violation, setting $g_L \equiv 0$.
Then, for the two-photon s-wave annihilation, we estimate the amplitude to be
\al{
\text{(i.a)}\quad \mathcal{M}_\text{s} \approx&   -\frac{i \alpha |g_R|^2 }{\pi} \frac{m_\chi^2}{m_\phi^2} [\hat{\mathbf{k}}\cdot (\boldsymbol{\epsilon}\times \boldsymbol{\epsilon}')] ,\nonumber      \\
& \text{for }\,  m_\phi \gg  m_\chi \gg m_f;  \label{s_ia}\\
\text{(i.b)}\quad \mathcal{M}_\text{s} \approx&   \frac{i  \alpha |g_R|^2 }{3\pi}\frac{m_\chi^4}{m_\phi^2 m_f^2}  [\hat{\mathbf{k}}\cdot (\boldsymbol{\epsilon}\times \boldsymbol{\epsilon}')],\nonumber      \\
& \text{for }\, m_\phi \gg m_f \gg m_\chi;\label{s_ib}\\
 \text{ (ii)}\quad \mathcal{M}_\text{s} \approx& \frac{i 2\alpha |g_R|^2 }{3\pi}  \frac{m_\chi^4}{m_f^4}  [\hat{\mathbf{k}}\cdot (\boldsymbol{\epsilon}\times \boldsymbol{\epsilon}')] ,\nonumber      \\
& \text{for }\,  m_f \gg m_\phi \gg m_\chi. \label{s_ii}
 }

As discussed in the previous section, the effective two-photon coupling to two Majorana fermions is described by a dimension-7 term in the Lagrangian so that one would anticipate that the annihilation amplitude would be suppressed by a factor of $\Lambda^{-3}$, for effective interaction scale $\Lambda$.  But, we see in the scenarios considered above, Eqs.~(\ref{s_ia}--\ref{s_ii}), a detailed computation of the amplitude reveals that the mass suppression of the amplitude is dictated, to a degree, by the relative mass of the charged fermion in the loop.  If this fermion mass is relatively small, then the amplitude is only suppressed by a factor of $\Lambda^{-2}= m_\phi^{-2}$. At the other extreme, if $m_f$ dominates all other masses, then the amplitude is suppressed by $\Lambda^{-4}= m_f^{-4}$.  As a result, the degree to which this s-wave process contributes to the total annihilation amplitude will vary accordingly.

\section{Relic density \label{sec_rd}}

The total annihilation cross section is the crucial particle-physics ingredient that determines the relic density of a thermal DM particle.  In the early universe the temperature is high enough so that the DM particles are not only in thermal equilibrium but also can be produced by SM particles through pair creation.  As the universe cools and expands the DM remains in thermal equilibrium, but its comoving number density decreases because SM particles no longer have sufficient energy to produce  DM.  Around the time that the DM annihilation rate roughly matches the universes's expansion rate, DM drops out of thermal equilibrium, and its comoving number density at this point essentially freezes out.  As a rule then, the weaker the annihilation cross section the greater the relic mass density of DM.  

For the total annihilation cross section, the s- and p-wave contributions sum incoherently so that $\sigma_\text{tot} = \sigma_\text{s} + \sigma_\text{p}$.  In the CoM frame, the differential cross section is $\frac{\diff \sigma}{\diff \Omega} |v_\text{rel}| = \frac{1}{2^7 \pi^2 m_\chi^2} |\mathcal{M}|^2$ so that for the s-wave process we have
\begin{equation}
\sigma_{\text{s}_0}:= \sigma_\text{s} |v_\text{rel}| =  \frac{1}{32 \pi m_\chi^2} |\mathcal{M}_\text{s}|^2 , 
\end{equation}
averaging over initial spin states and summing over the final-state polarizations.
If only one possible final state existed for the p-wave process, then this channel's contribution to the total cross section would be
\begin{equation}
\sigma_\text{p} |v_\text{rel}| =  \frac{2}{3}  \alpha  f_a^2   m_\chi^2  v_\text{rel}^2.
 \end{equation}
However, we must consider all final charged fermionic SM states which are kinematically accessible; i.e., all charged  fermions with masses less than $m_\chi$.  Because quarks have fractional charges $Qe$ and color degrees of freedom, we will account for this via a flavor factor $N_f = 3Q^2$. So, in what follows, the p-wave process will include the sum over all kinematically accessible SM fermionic final states. 
In computing the relic density, we must thermally average these contributions to the cross section.  Because the s-wave amplitude, $\mathcal{M}_\text{s}$, is velocity independent, the thermal average is trivial, but for the p-wave term, we have  $\langle \sigma_\text{p} |v_\text{rel}| \rangle = 4  \alpha  f_a^2   m_\chi T:=\sigma_{\text{p}_0} x^{-1} $ where $x:= m_\chi/T$ as in Ref.~\cite{anapole_dm1}.

Rather than solve the Boltzmann equation, we make use of the approximations, accurate to a few percent, in Refs.~\cite{Scherrer:1985zt,kolb_turner} that allow us to simply compute the relic DM mass density (relative to the critical density), $\Omega_\text{DM}$, and the freeze-out temperature, $T_f = m_\chi/x_f$.  First, we focus on the situation in which only the s- or p-wave process is relevant.  For this scenario, we write the thermally averaged annihilation cross section as $\langle \sigma |v_\text{rel}|\rangle = \sigma_0 x^{-n}$, where $n=0$ if the cross section is s-wave and $n=1$ for a p-wave process.  The DM relic density can be estimated via
\al{
\Omega_\text{DM}  =& \frac{3.79\,  s_0}{\rho_\text{crit}  g_*^{\frac{1}{2}} M_\text{Pl}} \frac{(n+1)x_f^{n+1} }{ \sigma_{0}} ,\\
x_f =& \log \left[  0.076\, g_*^{-\frac{1}{2}}(n+1) M_\text{Pl} m_\chi  \sigma_0   \right] \nonumber\\
& -\left(n+\frac{1}{2}\right)  \log[ \log[  0.076\, g_*^{-\frac{1}{2}}(n+1) M_\text{Pl} m_\chi \sigma_0 ]].
}
In these equations, $s_0$ represents the universe's present entropy density; $\rho_\text{crit}$ is the critical energy density; $g_*$ represents the relativistic degrees of freedom at the freeze-out temperature; and $M_\text{Pl}$ is the Planck mass.  We refer to Ref.~\cite{pdg2016} for values for the astrophysical parameters.

Before considering our explicit model calculations, it is instructive to estimate the impact that an s-wave cross section has on the DM density or properties vis-\`a-vis a p-wave dominated cross section.   
Suppose, for example, that $\sigma_{\text{s}_0} \sim \sigma_{\text{p}_0}$; this is the case with scenario (i.a)  of our model calculation in which $m_\phi \gg  m_\chi \gg m_f$. For this situation, the velocity suppressed p-wave annihilation will result in a greater relic density, $\Omega_\text{DM}^\text{p}$, than that for an s-wave dominated annihilation, $\Omega_\text{DM}^\text{s}$.  In particular, we find  $\Omega_\text{DM}^\text{p}/\Omega_\text{DM}^\text{s} \sim 2 x_f$.  For a thermal WIMP, freeze out occurs for values of $x_f$ that are $\mathcal{O}(10)$.  As a result, the p-wave process would result in a DM relic density that is at least a factor of 20 greater than that for an s-wave process.

Suppose instead that we wished to fix model DM parameters so as to reproduce the observed DM relic density. For some models, it is possible that $\sigma_{\text{s}_0}$ and $\sigma_{\text{p}_0}$ have identical parameter dependences.  In particular, for scenario (i.a) of our explicit model, we find $\sigma_{\text{s}_0} \sim \sigma_{\text{p}_0} \sim \alpha f_a^2 m_\chi^2$  with $f_a^2 \sim \alpha  |g_R|^4/m_\phi^4$.  For a given DM candidate mass $m_\chi$, if we were to use only the s- or p-wave annihilation cross section to estimate the anapole moment $f_a$, our results would differ significantly.  In particular, the value for $f_a$ determined by the p-wave process will be a factor of roughly $\sqrt{2x_f} \gtrsim 4$ greater than that determined by the s-wave process.

This is precisely what we will consider here. That is, for our three model scenarios, we will  determine what anapole moment, $f_a$, and DM mass, $m_\chi$, will yield the measured relic DM density $\Omega_\text{DM} h^2 = 0.1186$ \cite{pdg2016}.   We will consider both the s- and p-wave processes independently, but because both could be relevant, we must determine their joint impact on the relic density.  Returning to Refs.~\cite{Scherrer:1985zt,kolb_turner}, we find that if both $\sigma_\text{s}$ and $\sigma_\text{p}$ are appreciable then the relic DM density and freeze-out temperature can be estimated as
\al{
\Omega_\text{DM}  =& \frac{3.79\,  s_0 x_f}{\rho_\text{crit}  g_*^{\frac{1}{2}} M_\text{Pl}} \left( \sigma_{\text{s}_0} + \frac{1}{2 x_f} \sigma_{\text{p}_0} \right)^{-1} ,\\
x_f =& \log \left[ \xi  \left(\sigma_{\text{s}_0} + \frac{\sigma_{\text{p}_0}}{\log[\xi \sigma_{\text{s}_0}]} \right) \right]  -\frac{1}{2}  \log[ \log[  \xi \sigma_{\text{s}_0} ]],
}
with $\xi = 0.076\, g_*^{-\frac{1}{2}} M_\text{Pl} m_\chi  $.

\begin{figure}[h]
\includegraphics[width=6.5cm]{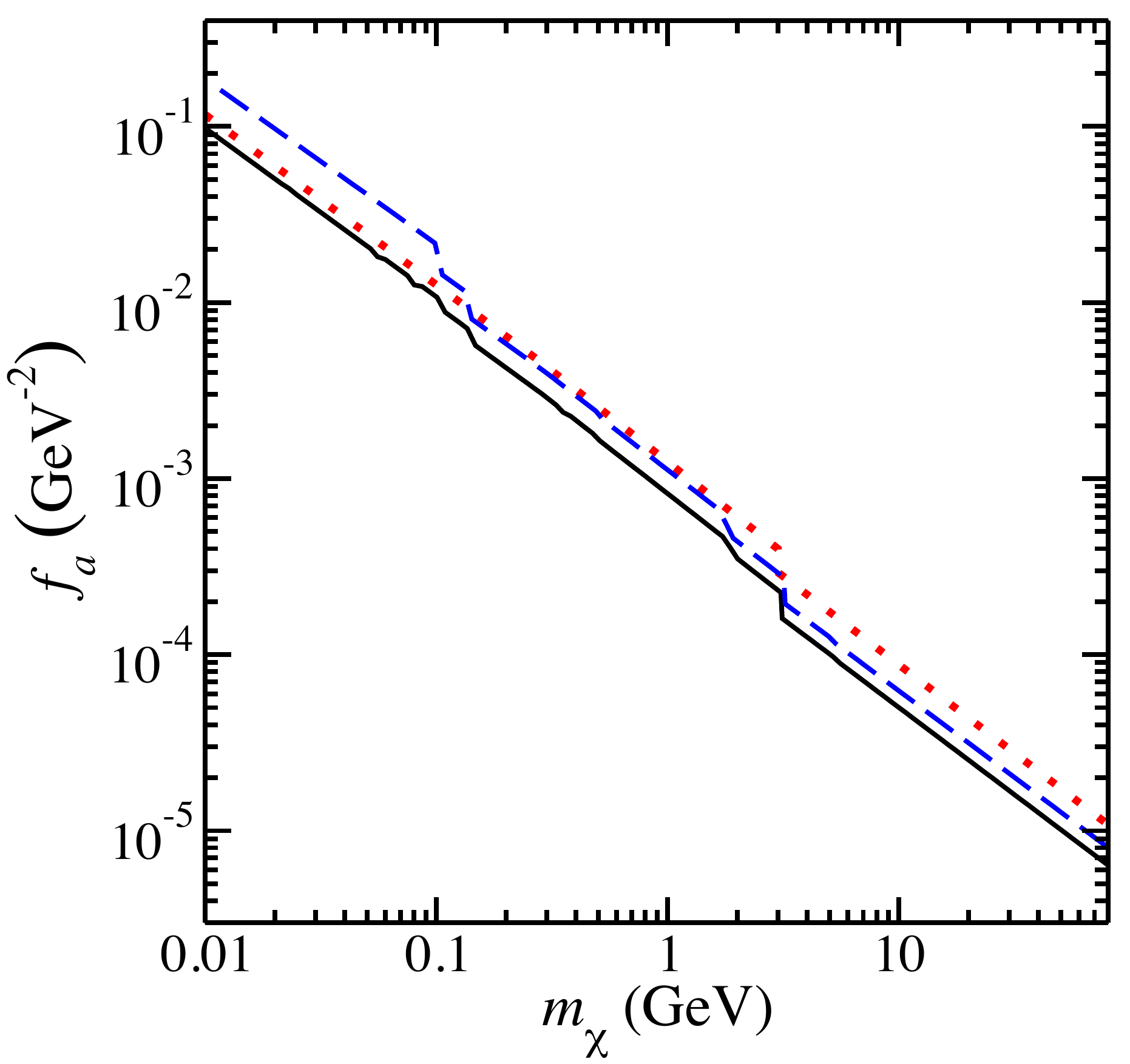}
\caption{(Color online) The anapole moment and DM mass which reproduce the observed relic DM density.  The dashed (blue) line employs only p-wave annihilation; the dotted (red) line employs only s-wave annihilation; and the solid (black) line employs both s- and p-wave processes. We assume $m_\phi = 10 m_\chi$ and $m_\chi = 10 m_f$.  \label{fig4}}
\end{figure}

We begin with case (i.a) with $m_\phi \gg m_\chi \gg m_f$.  Our model calculation involves four free parameters subject to only two constraints, so the system is underdetermined.  To enforce the assumed mass hierarchy, we set $m_\phi  = 10 m_\chi$ and $m_\chi = 10 m_f$.  Note, in our approximations the perturbative parameter was the ratio of the masses squared so this assumed hierarchy means that our neglected terms are $\mathcal{O}(10^{-2})$.  With these mass ratios fixed, we solve for $g_R$ and $m_\chi$ that yield the correct relic DM density.  Expressing this in terms of the DM mass and anapole moment, our results are in Fig.~\ref{fig4}.  The (blue) dashed curve employs only the p-wave annihilation into SM charged fermions; it closely reproduces the results in Ref.~\cite{anapole_dm1}.    The (red) dotted curve employs only the s-wave annihilation into photons, and the (black) solid curve uses both the s- and p-wave processes to fit the relic density. 
For lower DM masses, the s-wave term in the annihilation cross section dominates the determination of the DM relic density (and hence the anapole moment which achieves the observed relic density).  As the DM mass increases, more p-wave annihilation channels are accessible, and the importance of the p-wave cross section eclipses that of the s-wave process.  Still, for masses between a few GeV and 80 GeV, the s-wave process does contribute non-trivially to the determination of the anapole moment.  In this mass region, neglecting the s-wave process will result in a roughly 25\% greater value for $f_a$ than if both processes are considered.  The cross section is proportional to the square of the anapole moment so neglecting the s-wave process can impact the relic DM density at the level of 50\%.

For case (i.b), we consider  $m_\phi \gg m_f \gg m_\chi$.  Working with the same mass hierarchy scale, we set $m_\phi  = 10 m_f$ and $m_f = 10 m_\chi$.  Comparing the amplitude for (i.b), Eq.~(\ref{s_ib}), with that from (i.a), Eq.~(\ref{s_ia}), we find that the (i.b) amplitude is suppressed by an additional factor of $10^4$.  As such, the impact of the s-wave process for this case is negligible relative to the p-wave process so that the anapole moment $f_a$ is set almost exclusively by the p-wave process.   
For case (ii), the suppression of the s-wave process, Eq.~(\ref{s_ii}), is even greater.
Assuming a mass ratio of $m_f = 100 m_\chi$, the s-wave process will result in a value of $f_a$ around $10^6$ greater than that determined by the p-wave process.

\section{Confronting constraints on s-wave annihilation}

Assuming s-wave annihilation, the nominal cross section needed to reproduce the measured relic DM density is roughly $\langle \sigma |v_\text{rel}|\rangle \sim 3 \times 10^{-26}$ cm$^3$/s, providing an upper bound on the s-wave annihilation cross section considered herein. But, there are more stringent constraints on this quantity which could impact the three model scenarios considered above.
Energy injection from DM annihilation at the time of recombination can significantly impact cosmic microwave background (CMB) anisotropies. 
 From the Planck satellite's high precision measurements of the CMB  \cite{planck2015_cosmo},  it is shown that the s-wave cross section for annihilation into photons must be less than  $\langle \sigma |v_\text{rel}|\rangle \sim 10^{-29}$ cm$^3$/s at 0.01 GeV with the bound rising monotonically to $\sim 10^{-26}$ cm$^3$/s at 10 GeV \cite{slatyer_cmb}.  Observations from the Fermi Large Area Telescope (LAT) provide complementary constraints on this cross section through the non-observation of monoenergetic gamma rays, attributable to DM annihilation, from the Milky Way halo \cite{ ackermann_prd}.  The most severe constraints on the annihilation cross section come from region of interest R3. For a DM mass of 0.2 GeV one finds the upper bound $\langle \sigma |v_\text{rel}| \rangle \sim 6 \times10^{-31}$ cm$^3$/s  which rises, nearly monotonically, to a bound of $5 \times 10^{-29}$ cm$^3$/s at 80 GeV.  (We quote the median expected limit from the Monte Carlo simulations because it is smoother than the actual data.)  

We now examine the impact of these additional constraints for case (i.a) considered above with $m_\phi \gg m_\chi \gg m_f$ assumed.  Referring to Fig.~\ref{fig4}, we see that the anapole moment as determined by exclusive s-wave annihilation closely tracks that determined by the total cross section.  Because this anapole moment is chosen so as to reproduce the relic DM density,  we expect the s-wave annihilation cross section to be close to $3 \times 10^{-26}$ cm$^3$/s. This is what we find. Using the values for $f_a$ established by the total annihilation cross section,  the s-wave cross section falls between $1 \times 10^{-26}$ cm$^3$/s and $3 \times 10^{-26}$ cm$^3$/s for DM masses $m_\chi < 3$ GeV; beyond 3 GeV, the cross section drops to $6 \times 10^{-27}$ cm$^3$/s.  Comparing these values with the constraints from the Planck and Fermi LAT observations, we find that this model of anapole DM is not viable.  

For the remaining cases, (i.b) and (ii), the s-wave mode is suppressed relative to the p-wave mode, so it is this latter process that predominantly determines the value of the anapole moment that fits the observed relic DM density.  In both cases, this anapole moment is given by the dashed (blue) curve in Fig.~\ref{fig4}.  Focusing on case (i.b) with $m_\phi \gg m_f \gg m_\chi$, we see that the size of the anapole moment, Eq.~(\ref{ana_i}), is primarily set by the heavier mass $m_\phi$, though there is a logarithmic dependence upon $m_f$. 
Above, we worked with fixed mass ratios because our system was underdetermined.  Because we are incorporating the additional constraint on s-wave annihilation, we will relax these mass ratios somewhat.    We will still assume that $m_\phi = 10 m_f$ in the slowly varying logarithm so that $f_a$ determines the ratio $|g_R|/m_\phi$, but otherwise we allow $m_f$ to be a free parameter in the s-wave annihilation cross section.   Post hoc, we find that this assumption introduces negligible error.

\begin{figure}[h]
\includegraphics[width=6.5cm]{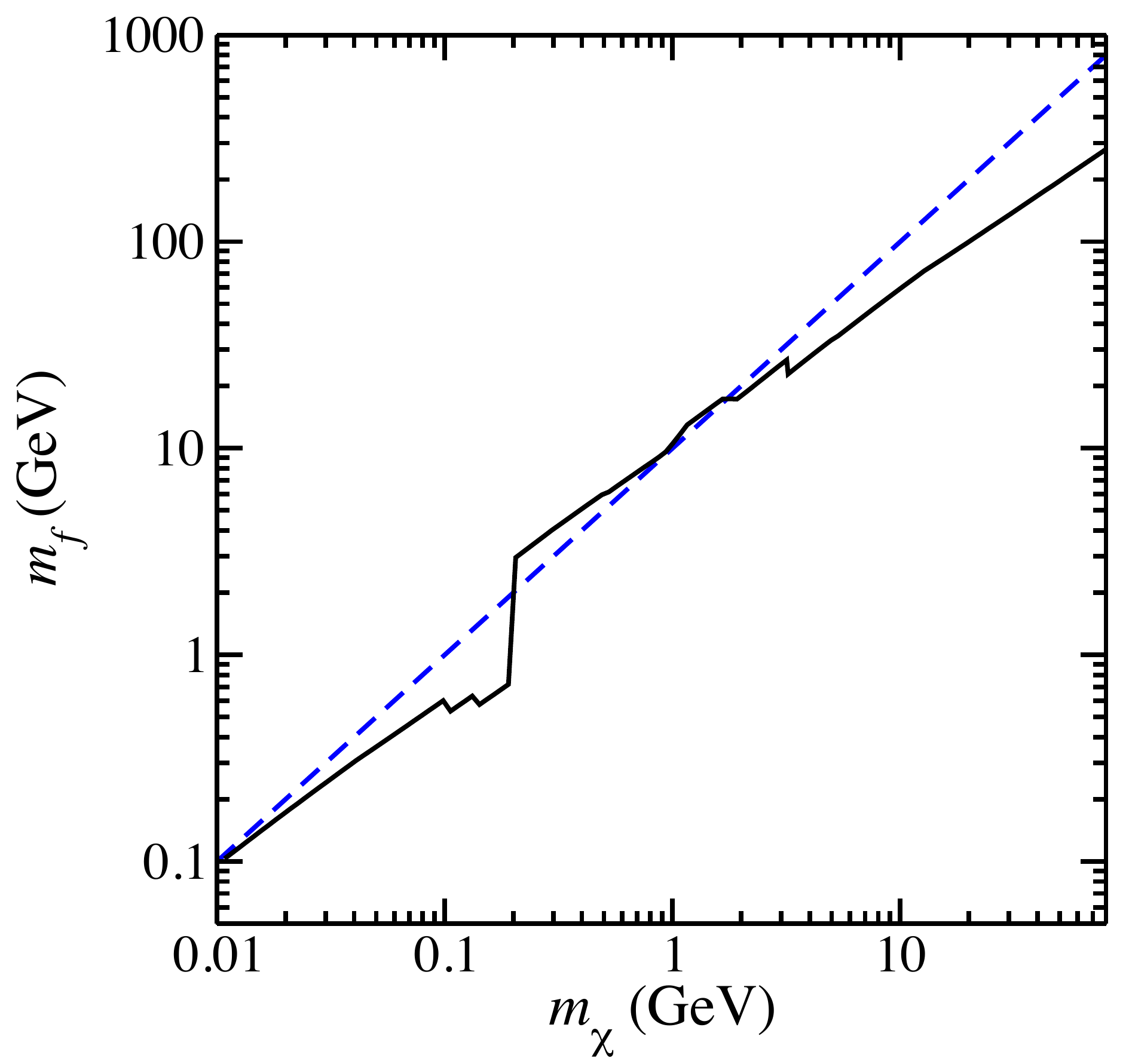}
\caption{(Color online)  The (black) solid curve shows the lower bound on $m_f$ for case (i.b) as set by the constraints on s-wave DM annihilation into photons.  The (blue) dashed curve shows the assumed mass ratio $m_f = 10 m_\chi$ used in Sec.~\ref{sec_rd}. \label{fig5}}
\end{figure}

For case (i.b), the (black) solid curve in Fig.~\ref{fig5} shows the lower bound on the mass $m_f$ as set by the Planck CMB \cite{slatyer_cmb} and Fermi LAT \cite{ackermann_prd} constraints on DM s-wave annihilation into photons. For DM masses below 0.2 GeV, we use the CMB constraint, but beyond that, we use the more stringent constraints from the Fermi LAT observations.  This accounts for the sizable discontinuity in the curve at that mass.  Additionally, we plot the (blue) dashed curve which depicts the assumed ratio $m_f = 10 m_\chi$ used in Sec.~\ref{sec_rd}.
We see that, for the most part, the mass ration $m_f = 10 m_\chi$ evades the constraints on s-wave annihilation.  Between 0.2 GeV and 1.0 GeV, one would need to set $m_f \sim 15 m_\chi$ in order to produce a viable model.  

\begin{figure}[h]
\includegraphics[width=6.5cm]{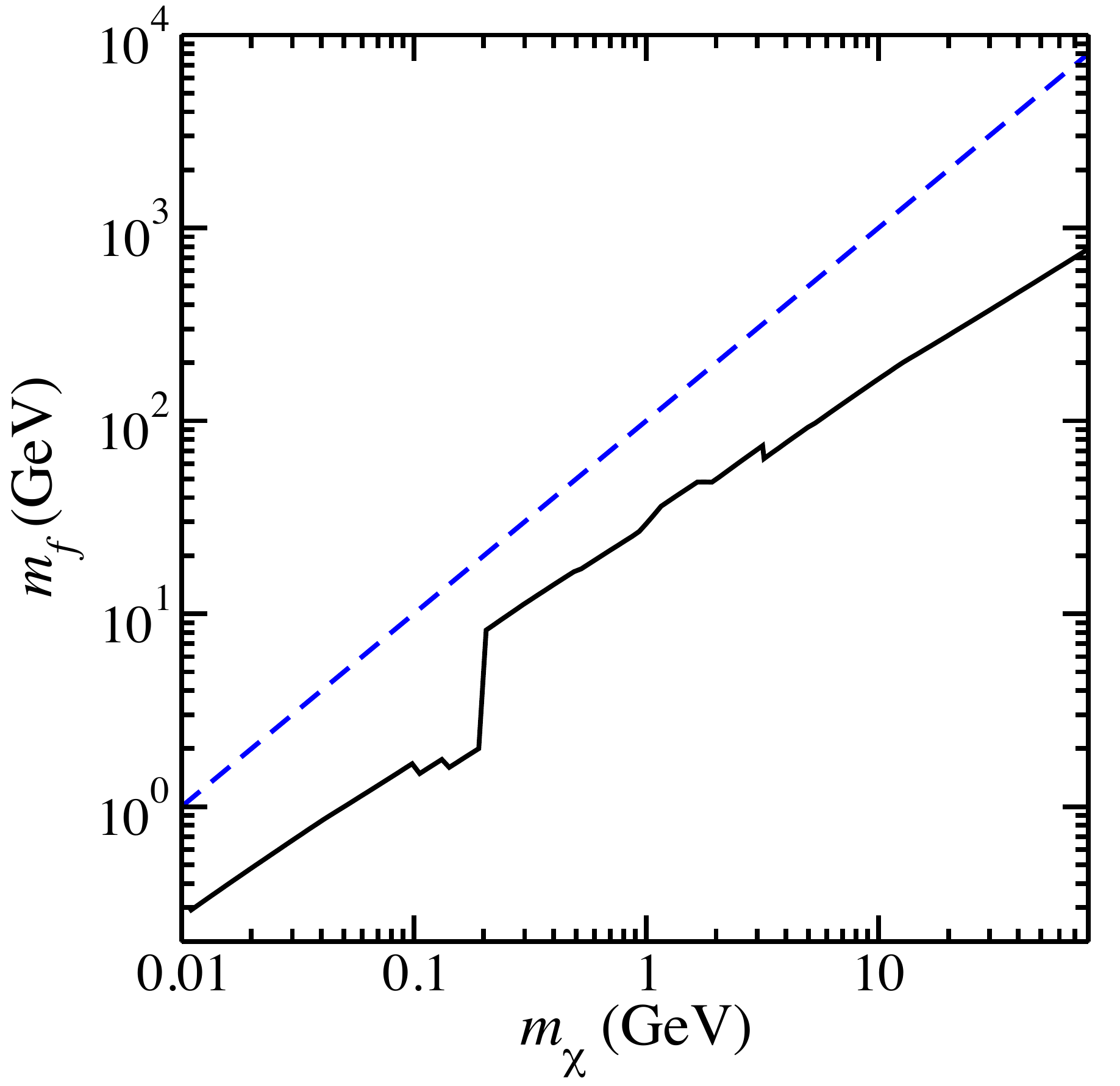}
\caption{(Color online)    The (black) solid curve shows the lower bound on $m_f$ for case (ii) as set by the constraints on s-wave DM annihilation into photons.  The (blue) dashed curve shows the assumed mass ratio $m_f = 100 m_\chi$ used in Sec.~\ref{sec_rd}.  \label{fig6}}
\end{figure}

We similarly implement the s-wave constraints for the mass hierarchy in case (ii).  In this scenario, the anapole moment, Eq.~(\ref{ana_ii}), is largely determined by the coupling $g_R$ and dominant mass $m_f$, though there is logarithmic dependence upon $m_\phi$.  As before, we fix $m_f= 10 m_\phi$ in the logarithm.  Then, we extract $|g_R|/m_f$ from the anapole moment that yields the observed relic DM density.  In applying the s-wave constraints, the dominant mass $m_f$ appears as a free parameter in the amplitude in Eq.~(\ref{s_ii}).  The resulting lower bound on $m_f$ is depicted as the (black) solid curve in Fig.~\ref{fig6}; additionally, we plot as the (blue) dashed curve the assumed mass ratio $m_f = 100 m_\chi$ used in Sec.~\ref{sec_rd}.  In comparing the two curves, we see that the mass ratios used in Sec.~\ref{sec_rd} result in a sufficiently small s-wave annihilation cross section so as to avoid the constraints from the Planck CMB and Fermi LAT observations. 

\section{Conclusion}

Because Majorana fermions are generally polarizable, these particles can annihilate into two photons in an s-wave process.  In a UV-complete model of anapole dark matter, this diphoton annihilation channel can rival the tree-level p-wave annihilation into charged SM fermions.  We explored this notion through an explicit  computation in which the anapole moment is generated via a one-loop process where the Majorana fermion couples to a charged scalar and fermion.  

We found that if the charged fermion is the lightest particle in the model, then the s-wave annihilation cross section was commensurate in size with the tree-level cross section.  In this case, the s-wave mode was important in setting the size of the anapole moment that would yield the correct relic DM density.  As such, when confronted with stringent astrophysical constraints on s-wave DM annihilation into photons, we find that this model is not viable.  
However, if the charged fermion's mass is greater than the  DM candidate's mass as in cases (i.b) and (ii), then the relic DM density is predominantly determined by the tree-level p-wave process.  For these two scenarios, we could easily evade the upper bound on the s-wave annihilation cross section.

As a caveat, we only considered DM masses up to 80 GeV.  Beyond this limit,  more annihilation channels open up for the tree-level process (namely, annihilation into $W$ bosons and top quarks).  As such, the p-wave process will increase in importance relative to the s-wave process, but we anticipate the the s-wave process still will be an important annihilation channel for case (i.a).

\section{Acknowledgment}
This work was funded, in part, by a Mellon Junior Sabbatical Fellowship from the University of Puget Sound.

\section{Appendix: Approximating the integrals $I_j$}

In this Appendix, we discuss how to approximate the integrals $I_j$, Eqs.~(\ref{I1}-\ref{I3}), that allow us to arrive at the approximate expressions for the s-wave annihilation amplitude into two photons. 
We assume the Majorana fermion is stable to decay so that $m_\chi < m_\phi+m_f$.  In this case, the integrals $I_2$ and $I_3$ are real; however, $I_1$ has an imaginary part whenever $m_f <m_\chi$ or $b<a$.  
All integrals can be expressed in terms of dilogarithms after factoring the quadratic expression in the arguments of the logarithms of each integrand.  We will focus on $I_2$ in detail, and integral $I_3$ readily follows by mapping $a \mapsto \frac{a}{b}$ and $b \mapsto \frac{1}{b}$, while $I_1$ requires closer examination. 

We factor the polynomial $a x^2 +(b-a - 1)x +1 = [A_+ x +1][A_- x +1]$ where 
\begin{equation}
A_\pm(a,b) :=  \frac{1}{2} \left[ (b-a-1) \pm \sqrt{(b-a-1)^2-4a}\right].
\end{equation}
Using the definition of the dilogarithm \cite{lewin}, we have
\al{
\int_0^1 \frac{1}{x} \log[a x^2 +(b-a - 1)x +1 ] \diff x =\nonumber
\\ - \mathrm{Li}_2 [-A_+(a,b)] - \mathrm{Li}_2 [-A_-(a,b)],
}
which results in the solution for $I_2$
\al{
I_2 =& - \mathrm{Li}_2 [-A_+(-a,b)] - \mathrm{Li}_2 [-A_-(-a,b)] \nonumber\\
&+ \mathrm{Li}_2 [-A_+(a,b)] + \mathrm{Li}_2 [-A_-(a,b)]. \label{i2_dilog}
}

For $I_2$, two limits are of interest:  (a) $m_\phi \gg  m_\chi, m_f$   and (b) $m_f \gg m_\phi, m_\chi$.  
For case (a), if the scalar mass dominates, then $a,b \ll 1$.  The leading order behaviors of the dilogarithm arguments are
\al{
A_+(a,b) \approx& -a -ab\\
A_-(a,b) \approx& -1+b +ab
}
omitting terms cubic in the small quantities.  Inserting these arguments into the dilogarithms in Eq.~(\ref{i2_dilog}), we find
\al{
I_2 \approx& 2a(1+b) \frac{\diff \phantom{x}}{\diff x} \mathrm{Li}_2 [x] |_{x=0} -2ab  \frac{\diff \phantom{x}}{\diff x} \mathrm{Li}_2[x] |_{x=1-b} \nonumber  \\
\approx & 2a(1 +b + b\log[b] ) ,
}
omitting terms cubic in the small quantities. We note the derivative of a dilogarithm is $\frac{\diff \phantom{x}}{\diff x} \mathrm{Li}_2 [x] =-\frac{1}{x} \log[1-x]$.

In case (b), where the charged fermion's mass dominates, we have  $\frac{a}{b}, \frac{1}{b} \ll 1$.   The arguments of the dilogarithms can be thus approximated
\al{
A_+(a,b) \approx& b-a-1-\frac{a}{b}-\frac{a}{b^2} -  \frac{a^2}{b^2},\\
A_-(a,b) \approx& \frac{a}{b}+ \frac{a}{b^2}+ \frac{a^2}{b^2};
}
so the integrals become
\al{
I_2 \approx&  2a\left(1+\frac{1}{b} + \frac{1}{b^2} \right)    \frac{\diff \phantom{x}}{\diff x} \mathrm{Li}_2 [x] |_{x=1-b} \nonumber \\
& - 2a\left(\frac{1}{b} + \frac{1}{b^2} \right) \frac{\diff \phantom{x}}{\diff x}  \mathrm{Li}_2 [x] |_{x=0}  \nonumber\\
\approx& \frac{2a}{b}\left\{ \log[b]-1+ \frac{1}{b}(2\log[b]-1)\right\} .
}

Turning to the integral, $I_1$, its integrand only directly involves masses $m_\chi$ and $m_f$. We will consider two limits:  $m_f \gg m_\chi$, relevant for cases (i.b) and (ii) above, and $m_\chi \gg m_f$, relevant for case (i.a) above.  As with the other integrals, we can factor the argument of the logarithm in the integrand of $I_1$ as $[B_+ x +1][B_-+1]$ with
\begin{equation}
B_\pm = 2\frac{a}{b} \left[ -1 \pm \sqrt{1-\frac{b}{a}} \right].
\end{equation}
The dilogarithms that occur in the solution $I_1 = -\mathrm{Li}_2[-B_+]-\mathrm{Li}_2[-B_-]$ can be simplified by use of the identity  \cite{lewin}
\begin{equation}
\mathrm{Li}_2[-z] +\mathrm{Li}\left[\frac{z}{1+z}\right] = -\frac{1}{2} \log^2[z+1],
\end{equation}
noting that $B_- = -\frac{B_+}{1+B_+}$.
As such, we have $I_1 = \frac{1}{2} \log^2[1+B_+]$.  

For $m_f > m_\chi$ or $b>a$,  $\log[1+B_+]$ is purely imaginary, and we can simplify the expression for $I_1$
\begin{equation}
I_1 = -2 \left(\tan^{-1}\frac{1}{\sqrt{ \frac{b}{a}-1 }}\right)^2.
\end{equation} 
Approximating this expression for large $m_f$, we find
\begin{equation}
 I_1 \approx  -2\frac{a}{b}- \frac{2}{3}\frac{a^2}{b^2}. 
\end{equation}

For $m_f < m_\chi$, the argument of the logarithm in $I_1$ is negative, resulting in an imaginary term
\begin{equation}
I_1 =\frac{1}{2} \log^2\left[ \frac{1+ \sqrt{1-\frac{b}{a}}}{1- \sqrt{1-\frac{b}{a}}}  \right] -\frac{1}{2}\pi^2 +i \pi \log\left[ \frac{1+ \sqrt{1-\frac{b}{a}}}{1- \sqrt{1-\frac{b}{a}}}  \right].
\end{equation}
If $m_f \ll m_\chi$, then the leading order contribution to the integral is
\begin{equation}
I_1\approx \frac{1}{2} \left( \log^2\left[4\frac{a}{b} \right] -\frac{b}{a} \log\left[4\frac{a}{b} \right]  \right)-\frac{\pi^2}{2} -i\pi  \log\left[4\frac{a}{b}\right].
\end{equation}

\bibliography{biblio}

\end{document}